\documentclass[prl,aps,twocolumn,showpacs,preprintnumbers,amsmath,amssymb,superscriptaddress]{revtex4-2}
\usepackage{dcolumn}
\usepackage{graphicx}
\usepackage[colorlinks=true,linkcolor=blue,citecolor=blue]{hyperref}
\usepackage{color}

\begin{document}

\title{Photo-induced Antiferromagnetic-ferromagnetic Transition and Electronic Structure Modulation in GdBaCo$_2$O$_{5.5}$ Thin Film}

\author{Yujun Zhang}
\email{zhangyujun@ihep.ac.cn}
\affiliation{Institute of High Energy Physics, Chinese Academy of Sciences, Yuquan Road 19B, Shijingshan District, Beijing, 100049, China}
\affiliation{Graduate School of Materials Science, University of Hyogo, 3-2-1 Kouto, Kamigori-cho, Ako-gun, Hyogo 678-1297, Japan}
\affiliation{Institute for Solid State Physics, University of Tokyo, 5-1-5 Kashiwanoha, Chiba 277-8581, Japan}
\author{Tsukasa Katayama}
\affiliation{Department of Chemistry, University of Tokyo, 7-3-1 Hongo, Bunkyo, Tokyo 113-0033, Japan}
\author{Akira Chikamatsu}
\affiliation{Department of Chemistry, University of Tokyo, 7-3-1 Hongo, Bunkyo, Tokyo 113-0033, Japan}
\author{Christian Sch\"u\ss ler-Langeheine}
\affiliation{Helmholtz-Zentrum Berlin f\"ur Materialien und Energie GmbH, Albert-Einstein-Stra\ss e 15, 12489 Berlin, Germany}
\author{Niko Pontius}
\affiliation{Helmholtz-Zentrum Berlin f\"ur Materialien und Energie GmbH, Albert-Einstein-Stra\ss e 15, 12489 Berlin, Germany}
\author{Yasuyuki Hirata}
\affiliation{Institute for Solid State Physics, University of Tokyo, 5-1-5 Kashiwanoha, Chiba 277-8581, Japan}
\author{Kou Takubo}
\affiliation{Institute for Solid State Physics, University of Tokyo, 5-1-5 Kashiwanoha, Chiba 277-8581, Japan}
\author{Kohei Yamagami}
\affiliation{Institute for Solid State Physics, University of Tokyo, 5-1-5 Kashiwanoha, Chiba 277-8581, Japan}
\author{Keisuke Ikeda}
\affiliation{Institute for Solid State Physics, University of Tokyo, 5-1-5 Kashiwanoha, Chiba 277-8581, Japan}
\author{Kohei Yamamoto}
\affiliation{Institute for Solid State Physics, University of Tokyo, 5-1-5 Kashiwanoha, Chiba 277-8581, Japan}
\author{Tetsuya Hasegawa}
\affiliation{Department of Chemistry, University of Tokyo, 7-3-1 Hongo, Bunkyo, Tokyo 113-0033, Japan}
\author{Hiroki Wadati}
\affiliation{Graduate School of Materials Science, University of Hyogo, 3-2-1 Kouto, Kamigori-cho, Ako-gun, Hyogo 678-1297, Japan}
\affiliation{Institute for Solid State Physics, University of Tokyo, 5-1-5 Kashiwanoha, Chiba 277-8581, Japan}

\begin{abstract}
We investigate both the ferromagnetic (FM) and antiferromagnetic (AFM) ultrafast dynamics of a strongly correlated oxide system, GdBaCo$_2$O$_{5.5}$ thin film, by time-resolved \mbox{x-ray} magnetic circular dichroism in reflectivity (XMCDR) and resonant magnetic \mbox{x-ray} diffraction (RMXD). A photo-induced AFM-FM transition characterized by an increase of the transient XMCDR beyond the unpumped value and a decay of RMXD was observed. The photon-energy dependence of the transient XMCDR and reflectivity can likely be interpreted as a concomitant photo-induced spin-state transition. 
\end{abstract}

\maketitle

Ultrafast magnetic dynamics and transient magnetic states induced by optical excitation have attracted a lot of research interest~\cite{1_Beaurepaire1996Ultrafast,2_Ogasawara2005General,3_Tsuyama2015Photoinduced,4_Takubo2017Capturing,5_Kirilyuk2010Ultrafast,6_Pontius2011Time,7_Chuang2013Real,8_Dean2016Ultrafast,9_Yamamoto2019Ultrafast,10_Yamamoto2019Element}.
With the rapid development of synchrotron \mbox{x-ray} sources with pulsed time structures, the role of probe in pump-probe experiments can be played not only by optical lasers but also by synchrotron \mbox{x-rays}~\cite{3_Tsuyama2015Photoinduced,4_Takubo2017Capturing,5_Kirilyuk2010Ultrafast,6_Pontius2011Time,7_Chuang2013Real,8_Dean2016Ultrafast} or \mbox{x-ray} free electron lasers~\cite{9_Yamamoto2019Ultrafast,10_Yamamoto2019Element}. Synchrotron \mbox{x-ray} can provide multiple probing techniques with element specificity, such as \mbox{x-ray} magnetic circular dichroism (XMCD)~\cite{3_Tsuyama2015Photoinduced,4_Takubo2017Capturing,9_Yamamoto2019Ultrafast}, resonant magneto-optic Kerr effect~\cite{10_Yamamoto2019Element} and resonant \mbox{x-ray} scattering~\cite{6_Pontius2011Time,7_Chuang2013Real,8_Dean2016Ultrafast}, etc. Capturing the element-specific dynamic behaviors of not only ferromagnetic~(FM) ordering, but also antiferromagnetic (AFM) ordering and complex magnetic structures become possible by employing synchrotron-based resonant magnetic \mbox{x-ray} diffraction (RMXD)~\cite{6_Pontius2011Time,7_Chuang2013Real,8_Dean2016Ultrafast}, which cannot be realized by optical laser probes. 

Materials with magnetic phase transitions are interesting from both views of fundamental science and practical application~\cite{11_Salamon2001The,12_Ward2009Elastically,13_Moruzzi1992Antiferromagnetic,14_Radu2010Laser}. 
One typical example is photo-induced AFM-FM transition observed in FeRh alloy~\cite{14_Radu2010Laser}. FeRh exhibits an AFM-FM transition at $\sim$370~K with increasing temperature~\cite{13_Moruzzi1992Antiferromagnetic} and laser-pumping can promote AFM FeRh to a transient FM state~\cite{14_Radu2010Laser}. However, due to the low photon energies of the magnetic sensitive Fe and Rh $L_{2,3}$ resonances, as well as the large $Q$-vector of the AFM ordering in FeRh~\cite{13_Moruzzi1992Antiferromagnetic}, it is technically unfeasible to investigate the dynamics of the AFM ordering even by using the synchrotron \mbox{x-ray} probes.

Recently, double perovskite $RE$BaCo$_2$O$_x$ ($5<x<6$, $RE$: rare earth elements) systems~\cite{15_Garcia2008Magnetic,16_Miao2017Hole,17_Katayama2018Ferromagnetism,18_Taskin2005Fast} have been intensely investigated due to their intriguing physical properties such as metal-insulator transition, AFM-FM transition, spin-state transition/ordering and high oxygen conductivity, etc. 
The oxygen concentration in these systems is often variable.
When $x\simeq$ 5.5, the oxygen vacancies can order into columns along $a$-axis, resulting in non-equivalent Co sites coordinated by either oxygen octahedra or oxygen pyramids~\cite{15_Garcia2008Magnetic,16_Miao2017Hole,17_Katayama2018Ferromagnetism,19_hu2012spin}. The nominal valence state of Co is 3+ and the competition between crystal field splitting and Hund coupling can lead to various spin-states
~\cite{15_Garcia2008Magnetic,16_Miao2017Hole,19_hu2012spin,20_Roy2002Observation,21_Vogt2003Pressure,22_Kimura2008Field,23_Yokoyama2018Tensile}. The spin-state of Co$^{3+}$ is sensitive to 
temperature, pressure~\cite{21_Vogt2003Pressure}, magnetic field~\cite{22_Kimura2008Field}, epitaxial strain~\cite{23_Yokoyama2018Tensile} and lattice distortions~\cite{15_Garcia2008Magnetic,16_Miao2017Hole}, etc.

AFM-FM transition coupled with spin-state transition (SST) were observed in some systems like GdBaCo$_2$O$_{5.5-x}$~\cite{15_Garcia2008Magnetic,19_hu2012spin}. Antiferromagnetism with an order vector of (0~0~0.5) below $\sim$230~K was detected in GdBaCo$_2$O$_{5.5-x}$ by RMXD thanks to its small AFM $Q$-vector~\cite{15_Garcia2008Magnetic}, providing the possibility to investigate the ultrafast dynamics of both the FM and AFM ordering simultaneously. In this work we investigate the photo-induced FM and AFM dynamics of GdBaCo$_2$O$_{5.5}$ (GBCO) thin film by time-resolved (TR) XMCD in reflectivity (XMCDR) and TR-RMXD at Co $L$ resonance within one experimental system. Photo-induced enhancement of XMCDR and decay of the RMXD were clearly observed, indicating a photo-induced AFM-FM transition. Photon energy dependence of the transient XMCDR and reflectivity suggests the possibility of a concomitant photo-induced SST. Both the electronic state change and the competition between AFM and FM exchange interactions may simultaneously contribute to the observed magnetic dynamics.

\begin{figure}[t]
	\includegraphics[width=\columnwidth]{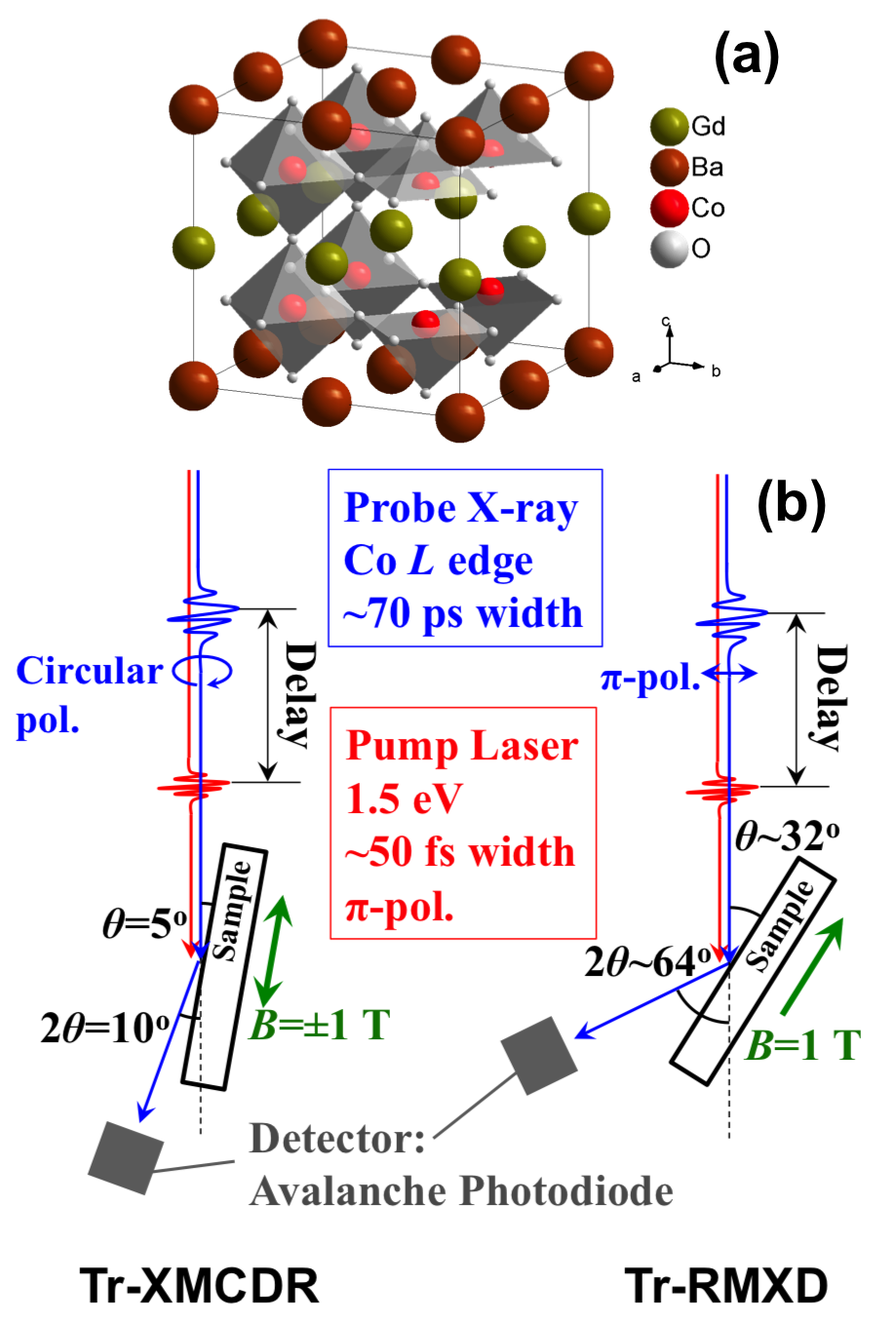}
	\caption{(a) Crystal structure of GBCO. (b) Setups of the TR-XMCDR and TR-RMXD experiments. 
		\label{fig1}
	}
\end{figure}

The 35-nm thick epitaxial GBCO thin film ($a$=$b$=7.81\AA, $c$=7.529\AA, the unit cell shown in Fig.~\ref{fig1}(a)) was grown on SrTiO$_3$(001) substrate by pulsed laser deposition and the detailed sample characterizations are reported elsewhere~\cite{24_Katayama2021}. Note that our GBCO film has  finite magnetization in the AFM phase, which should be attributed to the competition between AFM and FM exchange interactions, either the canting of AFM  moments~\cite{15_Garcia2008Magnetic} or the coexistence of AFM and FM phases~\cite{25_Wu2003Glassy,26_Hoch2004Evolution}, as explained in detail in the Supplemental Material~\cite{27_Supplemental}. The TR-XMCDR and TR-RMXD measurements were performed at beamline UE56/1-ZPM (FEMTOSPEX) of BESSYII by using setups shown in Fig.~\ref{fig1}(b). For TR-XMCDR measurements, \mbox{x-ray} with fixed circular polarization at the Co $L$ edge was used and an in-plane magnetic field of $\pm$1~T was switched to observe the magnetic contrast of the reflectivity. The incident angle was fixed at 5\textsuperscript{o}. The reflectivity was detected with an avalanche photodiode and boxcar integrated. For TR-RMXD measurements, we used horizontal linearly polarized \mbox{x-ray} ($\pi$-polarization) instead to enhance the magnetic diffraction, with a constant in-plane magnetic field of 1~T applied. The diffraction signal was integrated via photon counting. A Ti:sapphire laser ($\lambda$=800~nm, $\pi$-polarization, 3~kHz, pulse width $\sim$50~fs) was employed as the pump source. The spot sizes (horizontal$\times$vertical) of the pump laser and the probe \mbox{x-ray} were around 0.34$\times$0.38~mm$^2$ and 0.12$\times$0.04~mm$^2$, respectively. The time resolution of the measurements was limited to $\sim$70~ps by the pulse width of the probe \mbox{x-ray}. The pumped and unpumped signals were collected alternatively by recording the contributions from the pumped and unpumped bunches. The sample temperature was controlled by a liquid N$_2$ flow cryostat. Since the measurement geometries of XMCDR and RMXD were different, the surface-transmitted laser fluence was calibrated by measuring the incident-angle dependence of the laser power reflected by the sample surface. All the laser fluences mentioned below are calibrated fluences which are absorbed by the sample. 

\begin{figure}[t]
	\includegraphics[width=\columnwidth]{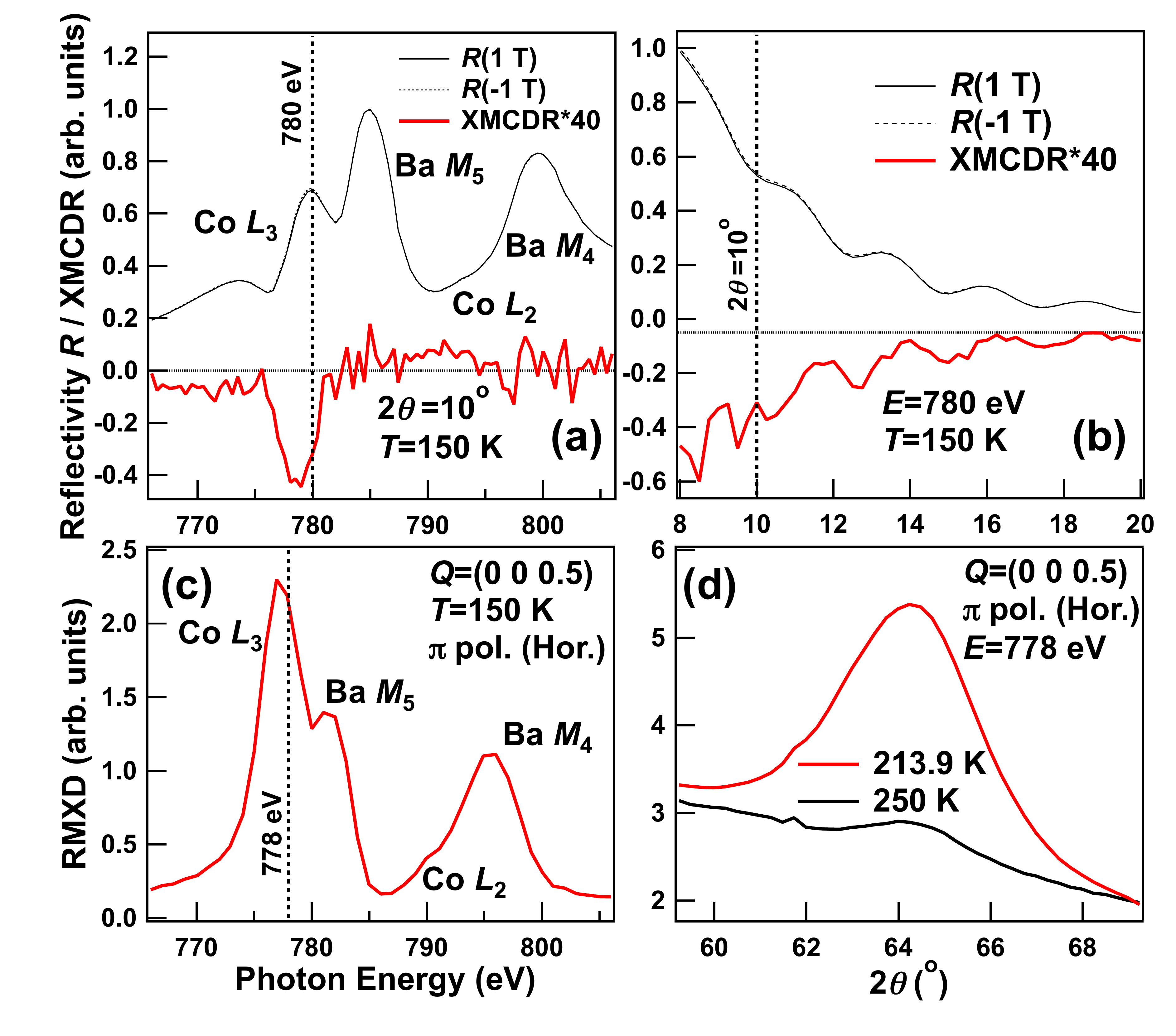}
	\caption{Photon energy scan (a) and $\theta$-2$\theta$ scan (b) of the static XMCDR. Photon energy scan (c) and temperature-dependent $\theta$-2$\theta$ scan (d) of the static (0~0~0.5) RMXD.
		\label{fig2}
	}
\end{figure}

Fig.~\ref{fig2}(a,b) show the static XMCDR of the GBCO thin film. Reflecticity $R$ and XMCDR are defined as $R$=($R_++R_-$)/2 and XMCDR=($R_+-R_-$)/2, respectively, where $R_+$ and $R_-$ are reflectivity measured by applying positive/negative magnetic fields. In the photon energy scan (Fig.~\ref{fig2}(a)), it can be observed that Co $L_3$ resonance is located near Ba $M_5$ resonance and the XMCDR mainly appeared in the range from 775~eV to 782~eV, which represents the Co magnetization. 
The thickness fringes in $\theta$-2$\theta$ reflectivity curve at the Co $L_3$ resonance (Fig.~\ref{fig2}(b)) is consistent with the nominal film thickness. An incident angle of $\theta$=5\textsuperscript{o} were chosen as the optimum detection geometry for the following TR-XMCDR measurements. Static RMXD results of the AFM ordering are shown in Fig.~\ref{fig2}(c,d). Photon energy scan confirmed the resonant behavior of the RMXD peak. Temperature-dependent $\theta$-2$\theta$ scans confirm that the RMXD peak vanishes above 230~K and corresponds to the magnetic diffraction of the (0~0~0.5) AFM ordering, which is consistent with the previous report~\cite{15_Garcia2008Magnetic}. The correlation length of the AFM ordering was estimated as $\sim$30.5~nm by the peak width~\cite{28_Yamamoto2018Thickness} in Fig.~\ref{fig2}(d), which is a little smaller than the film thickness possibly because of finite x-ray penetration depth or imperfections at the surface and interface. Photon energy was chosen at 778~eV for the TR-RMXD measurements.

\begin{figure}[t]
	\includegraphics[width=\columnwidth]{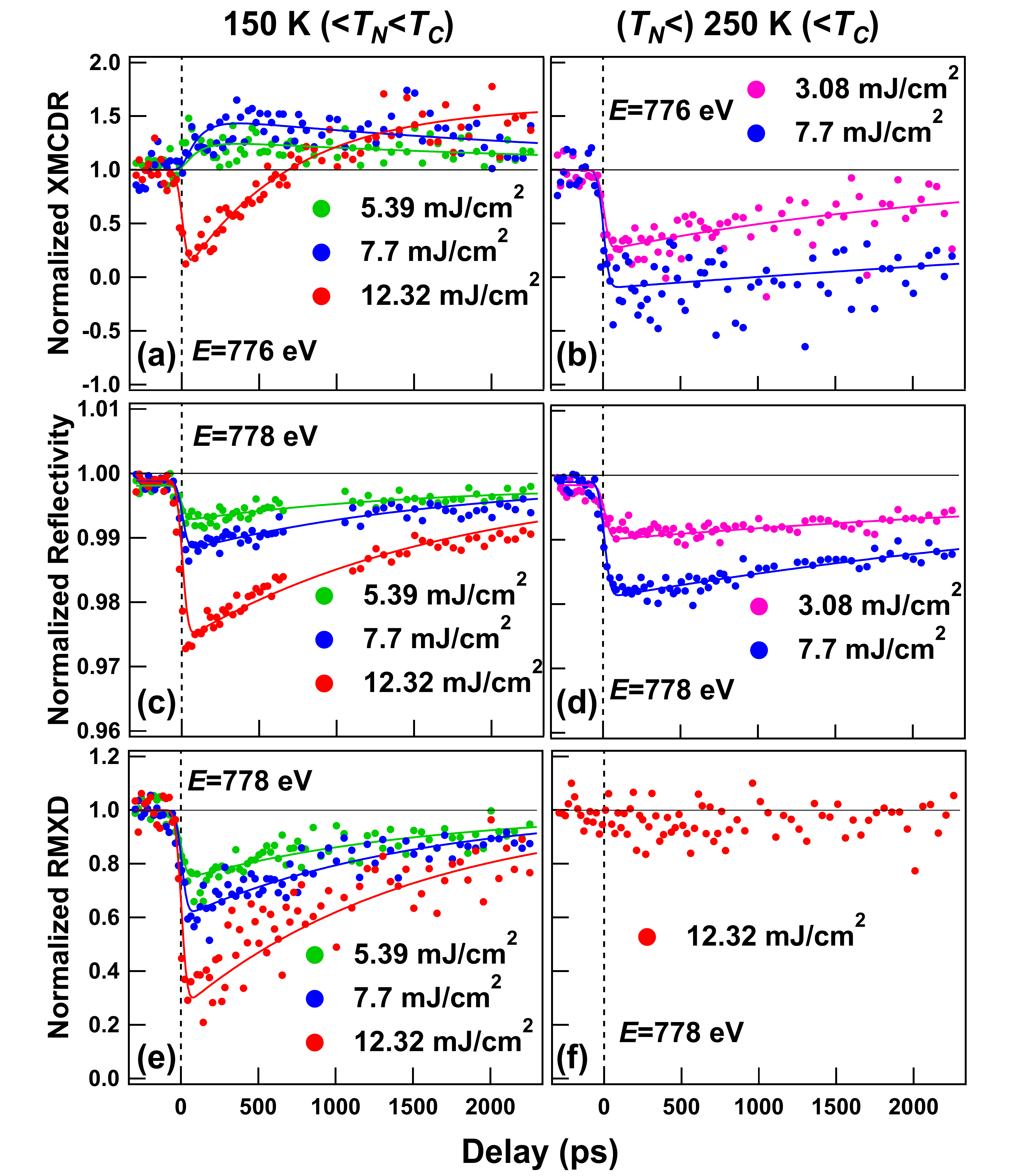}
	\caption{Delay scans of TR-XMCDR ((a) 150~K; (b) 250~K), TR-reflectivity ((c) 150~K; (d) 250~K) and TR-RMXD ((e) 150~K; (f) 250~K). All signals shown are normalized to the unpumped values. The curves are fittings of the data by Eq. (1). Before normalizing to the unpumped value, the RMXD in (e) is processed by first subtracting the reflectivity background and then taking the square root so that the signal is proportional to the AFM order parameter. The reflectivity background is calculated from the 250~K curve in Fig.~\ref{fig2}(d), which should contain no RMXD signal. The RMXD in (f) is only normalized to the unpumped signal without subtracting the reflectivity background and taking square root.
				\label{fig3}
	}
\end{figure}

Since GBCO exhibits a AFM-FM transition with low-temperature AFM phase~($T_N\sim$~230~K) and high-temperature FM phase~($T_C\sim$~300~K)~\cite{15_Garcia2008Magnetic,19_hu2012spin}, it is expected that laser pumping will promote a transient FM state at temperatures below the AFM-FM transition, as observed in FeRh~($T_N\sim$~375~K, $T_C\sim$~680~K)~\cite{14_Radu2010Laser}. This prediction was confirmed by our TR-XMCDR measurements shown in Fig.~\ref{fig3}(a). At 150~K, the transient increase of XMCDR beyond the unpumped value can be clearly observed. At lower laser fluences, the XMCDR increases immediately after the laser illumination and reaches a maximum of $\sim$150\% of the unpumped XMCDR. At higher laser fluence, the XMCDR exhibits a fast demagnetization followed by a increase of XMCDR beyond the unpumped value. Note that at 250~K (Fig.~\ref{fig3}(b)), which is above the AFM-FM transition temperature, the increase of the XMCDR beyond the unpumped value was not observed, rather a normal fast demagnetization and a slow recovery appeared. This confirms the occurrence of a photo-induced AFM-FM transition in GBCO thin film. Interestingly, the reflectivity of the sample also exhibits significant decay-recovery behavior (Fig.~\ref{fig3}(c,d)), inferring the concomitant change of electronic structure
~\cite{3_Tsuyama2015Photoinduced}. On the other hand, the TR-RMXD results in Fig.~\ref{fig3}(e) indicate that the (0~0~0.5) AFM peak exhibits a normal decay-recovery process. Above the AFM-FM transition temperature (250 K, Fig.~\ref{fig3}(f)), since the AFM peak does not exist anymore, no pump effect was observed.

Thus, it can be understood from the results in Fig.~\ref{fig3} that when the laser fluence is low, photo-induced AFM-FM transition occurs with the increase of net magnetization characterized by XMCDR and the decrease of the anti-parallel aligned component of Co moments characterized by the RMXD. When the laser fluence is as large as 12.32~mJ/cm$^2$, the fast decay of both XMCDR and RMXD should be attributed to a photo-induced demagnetization. During the recovery process, GBCO first enters a more FM transient state and the XMCDR increases to a value above the unpumped value after $\sim$1000~ps, as shown in Fig.~\ref{fig3}(a).

The delay scans in Fig.~\ref{fig3} are fitted by a double-exponential function,
$$I(t)=I_1 {\rm exp} (-t/\tau_{\rm dec.(enh.)})+I_2[1-{\rm exp} (-t/\tau_{\rm rec.})]\ (1)$$
convolved with a 70-ps-wide Gaussian time-resolution function to estimate the time scales of the decay/enhancement and recovery processes. The decay/enhancement processes are faster than the time resolution, while the recovery processes have different time scales for different signals. For instance, at 150~K and a laser fluence of 7.7~mJ/cm$^2$, the fitted recovery time scales are around 3.3~ns, 1.7~ns, and 1.5~ns for XMCDR, reflectivity and RMXD, respectively. The recovery time of the XMCDR is clearly different from that of the reflectivity and RMXD. This point will be explained in the following parts.

\begin{figure}[t]
	\includegraphics[width=\columnwidth]{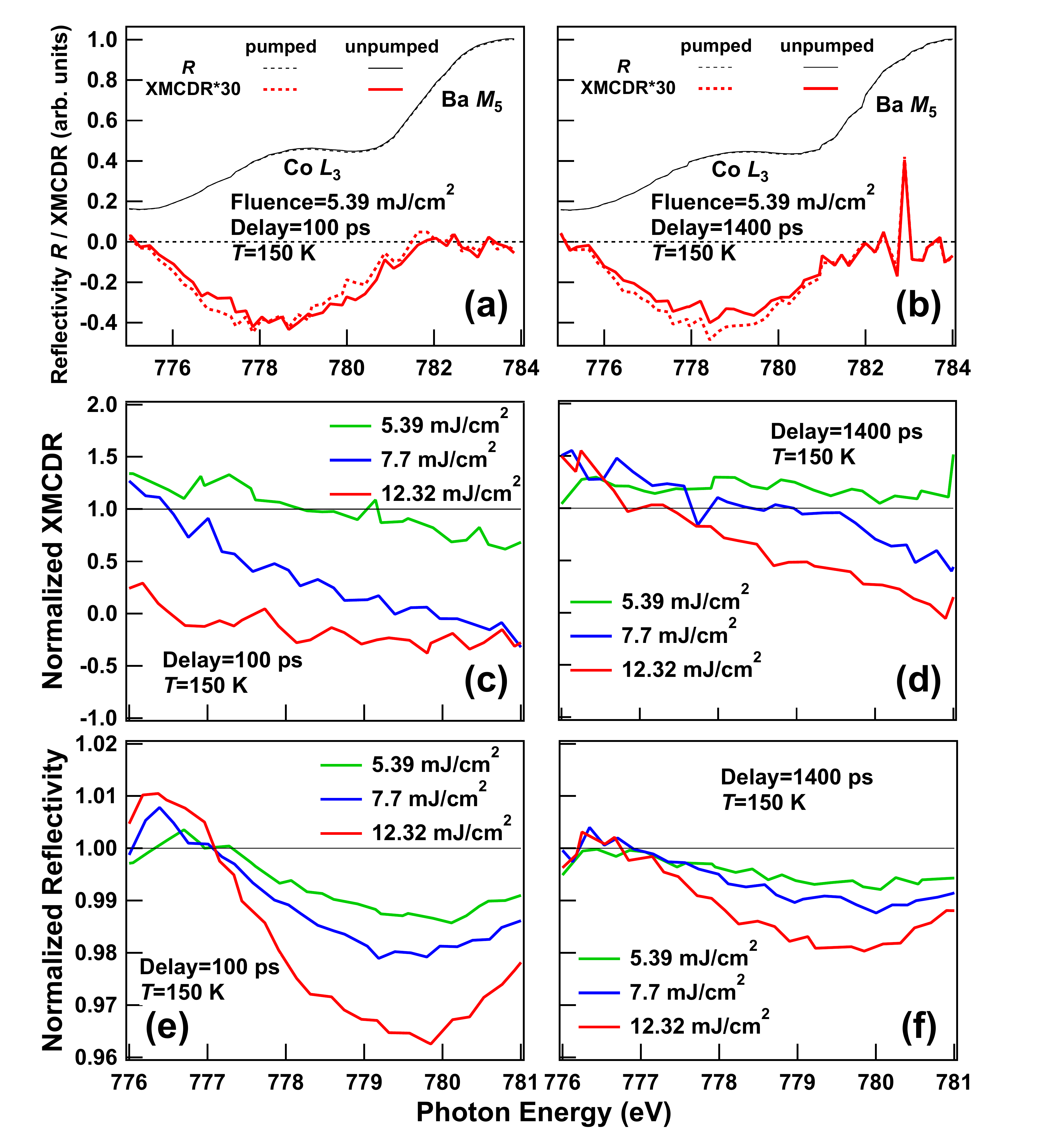}
	\caption{Energy dependence of the pump effect on reflectivity and XMCDR ((a) Delay=100~ps; (b) Delay=1400~ps). Normalized XMCDR and reflectivity spectra at different delay time ((c) XMCDR, 100~ps; (d) XMCDR, 1400~ps; (e) Reflectivity, 100~ps; (f) Reflectivity, 1400~ps). Curves in (c-f) are normalized to the corresponding unpumped values.
		\label{fig4}
	}
\end{figure}

To achieve a better understanding about the dynamic behaviors described above, we detailedly investigated the photon energy dependence of the transient XMCDR and reflectivity. Fig.~\ref{fig4}(a,b) show transient reflectivity and XMCDR spectra at a moderate laser fluence of 5.39~mJ/cm$^2$. At delay=100~ps~(Fig.~\ref{fig4}(a)), the pump effect is strongly photon energy dependent. In the energy range from 775~eV to 778~eV, the absolute value of XMCDR is enhanced by the pump laser. While in the energy range from 778~eV to 782~eV, the pump laser does the opposite effect. In contrast, at delay=1400~ps~(Fig.~\ref{fig4}(b)), the absolute value of XMCDR is enhanced by the laser-pumping in the whole energy range of Co $L_3$ resonance. Laser-fluence dependence of the pump effect on the XMCDR~(normalized to the corresponding unpumped values) can be more clearly observed in Fig.~\ref{fig4}(c,d). When the laser fluence is small~(green curves), a spectral weight transfer of XMCDR from higher energy to lower energy can be observed at delay=100~ps and then the spectrum evolves into an overall positive pumping effect of XMCDR within the range of the Co $L_3$ edge at delay=1400~ps. At higher laser fluences~(red curves), the increase of XMCDR is postponed by the demagnetization process. The spectral weight transfer of XMCDR also infers a photo-induced electronic state transition. Delay-time-dependent transient reflectivity spectra shown in Fig.~\ref{fig4}(e,f) can provide more insight. The pump effect of the reflectivity spectra is also strongly photon energy dependent. Pump laser causes a increase of reflectivity in the energy range from 776~eV to 777~eV and a decrease in the energy range near the resonance peak from 777~eV to 781~eV. 

In GBCO, the Co ions at the pyramidal sites are in stable high-spin~(HS) state and the Co ions at the octahedral sites are near the boarder of SST~\cite{19_hu2012spin}, so it is reasonable to predict that laser pumping can affect the spin-state of Co$^{3+}$ at the octahedral sites. Since the existence of intermediate-spin~(IS) state is still controversial in cobaltates, for simplicity we do not consider IS or just consider IS as an intermediate state between HS and low-spin~(LS) in the following discussion. It is previously reported that in Co$^{3+}$ perovskite oxides, when the population of HS octahedral Co$^{3+}$ increases, the intensity at the peak of the Co $L_3$ resonance decreases and a smaller feature increases at the lower energy side~\cite{15_Garcia2008Magnetic,19_hu2012spin}, which is consistent with the photo-induced change of transient reflectivity spectra reported here. At delay=100~ps, the similar direction of spectral weight transfer of XMCDR and reflectivity can be understood that HS Co$^{3+}$ contributes more to the magnetization due to its larger magnetic moment. At delay=1400~ps, the XMCDR spectral weight shifts back to the unpumped position due to the recovery of spin-state at octahedral sites. Thus, it is likely that a photo-induced LS-HS transition at the octahedral Co$^{3+}$ sites, as also reported previously in LaCoO$_3$~\cite{29_Izquierdo2014Laser,30_Izquierdo2019Monitoring}, occurs together with the AFM-FM transition. It is worth mentioning that photo-induced SST could not be predicated only by judging the spectral change of the reflectivity. There are other possible reasons for such spectral change, e.g. valence change~\cite{31_Guillou2013Coupled} or charge disproportionation~\cite{29_Izquierdo2014Laser} of Co ions. But in GBCO we conclude that photo-induced SST is a most probable explanation for the observed spectral change. The photo-induced transitions in GBCO significantly differ from the temperature-triggered AFM-FM transitions and SSTs in similar perovskite cobaltates in the sense that transitions of both magnetism and electronic structure are involved simultaneously. Generally the AFM-FM transition and SST are separated by the difference of transition temperatures in cobaltates~\cite{19_hu2012spin}. 

\begin{figure}[t]
	\includegraphics[width=\columnwidth]{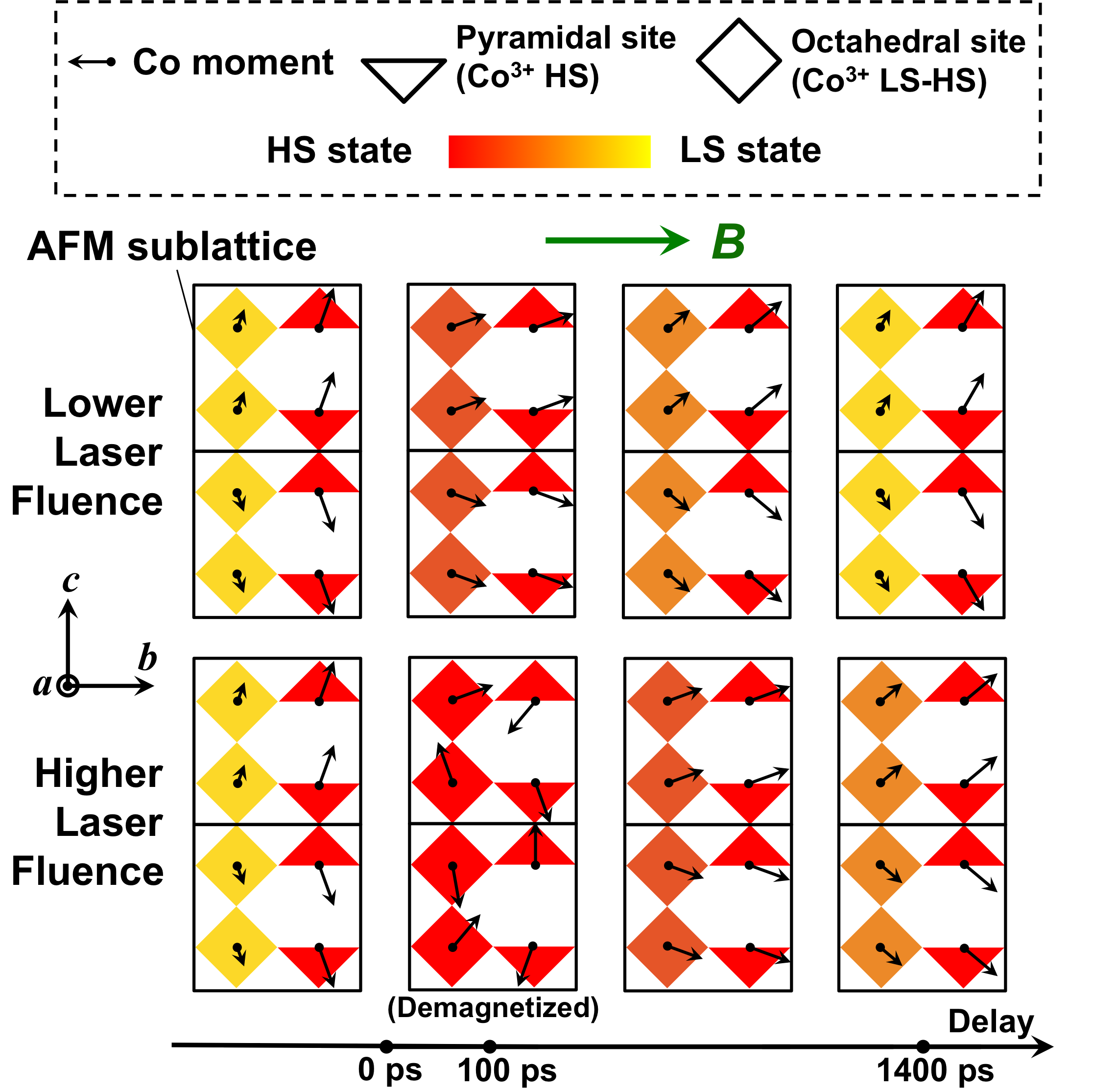}
	\caption{Schematic of the photo-induced ultrafast dynamics of magnetism and electronic structure in GBCO. 
		\label{fig5}
	}
\end{figure}

Now we can comprehensively describe the photo-induced dynamic behaviors of GBCO, as schematically shown in Fig.~\ref{fig5}. Laser illumination significantly changes the electronic structure of GBCO thin film, which can be characterized by the reflectivity. At the same time, pump laser also induces a AFM-FM transition, which can be characterized by the increase of XMCDR and decrease of the RMXD. The processes related to the magnetic degree of freedom can be interpreted as the change of either the AFM canting angle or the volume ratio between FM and AFM phases~\cite{27_Supplemental}. In Fig.~\ref{fig5} we are using the picture of AFM sublattice moment canting and the discussion should be analogous in the picture of phase separation. When the laser fluence is low, the system is first pumped into a FM state with higher fraction of HS octahedral Co$^{3+}$. The enhancement of XMCDR should be attributed to both the more parallel alignment of Co moments and the change of individual Co$^{3+}$ moment size due to the SST. Then the magnetism and spin-state relax with comparable speed. When a simple double exponential fitting is used, the contribution from both origins makes the fitted recovery time scale of XMCDR much slower than the recovery time scales of either reflectivity or RMXD. Meanwhile, the noise level of the XMCDR data cannot support meaningful fittings with more than two exponential functions. At high laser fluences, first the system is demagnetized. After the recovery from the demagnetized state, the system enters a more FM transient state with dominating HS Co$^{3+}$ population before further recovering to the original AFM state with mainly LS Co$^{3+}$ at octahedral sites.
 
To conclude, in this work we investigated the ultrafast dynamics of both FM and AFM ordering as well as the electronic structure in GBCO thin film. A photo-induced AFM-FM transition coupled with electronic structure transition, which is likely a SST at octahedral Co$^{3+}$ sites, was observed. The simultaneous contribution of moment alignment and spin-state change to the net magnetization in the transient state gives rise to the dynamic behaviors of XMCDR, reflectivity and RMXD. Our investigation presents a photo-induced AFM-FM transition in a strongly correlated oxide system and a new category of photo-induced phenomena involving both magnetic and electronic-structure degrees of freedom. The results indicate significant coupling of magnetism and electronic structure in strongly correlated systems, not only in static states but also in transient states and dynamic behaviors.

This work was supported by MEXT Quantum Leap Flagship Program (MEXT Q-LEAP) Grant No. JPMXS0118068681 and JSPS KAKENHI Grant No. 19H05824, 19H02594 and 17F17327. We thank HZB for the allocation of synchrotron radiation beamtime, Karsten Holldack and Rolf Mitzner for experimental support, and the supports provided by Nippon Sheet Glass Foundation for Materials Science and Engineering. Y. Z. thanks the supports from Natural Science Foundation of China (Grant No. 52002370) and Basic Research Funding of IHEP (Grant No. Y9515560U1).

\bibliographystyle{unsrt}

\end{document}